\documentstyle[12pt,psfig]{article}

\begin{document}
\title{Analysis of Kaon Production at SIS Energies\footnote{supported
by GSI Darmstadt, BMBF and DFG Graduiertenkolleg}}
\author{E. L. Bratkovskaya, W. Cassing and  U. Mosel \\
Institut f\"{u}r Theoretische Physik, Universit\"{a}t Giessen\\
35392 Giessen, Germany}
\date{}
\maketitle

\begin{abstract}
We analyse the production and propagation of pions and kaons in
heavy-ion reactions from 0.8 -- 1.8~A$\cdot$GeV  within a coupled channel
transport approach including the kaon production
channels $BB \to K^+YN, \ \pi B\to
K^+Y, \ BB \to NN K \bar{K}, \ \pi B\to N K\bar{K}, \ K^+B\to K^+B $ \
and $\ \pi \pi\to K \bar{K}$. Assuming the hyperon selfenergy to be 2/3
of the nucleon selfenergy we find that  all inclusive experimental
$K^+$ spectra at SIS energies can be reproduced reasonably well
without any selfenergies for the kaons although a slightly repulsive kaon
potential cannot be excluded by the present data on kaon spectra and flow.
\end{abstract}
\vspace{1cm}



\newpage
\section{Introduction}
The study of hadron properties in the dense nuclear medium via
relativistic nucleus-nucleus collisions is a major aim of high energy
heavy-ion physics. Especially the production of particles at
'subtreshold' energies is expected to provide some valuable insight
into the properties of hadrons at high baryon density and temperature
\cite{Cass}. In this respect selfenergy effects in the production of
particles have been found previously for antiprotons by a number of
groups~\cite{Teis}--\cite{Batko95} though the actual magnitude of the
attractive $\bar{p}$-potential in the nuclear medium is still a matter
of debate. Also antikaons according to Refs.~\cite{GB1}--\cite{waas}
should feel strong attractive forces in the medium whereas the kaon
potential is expected to be slightly repulsive at finite baryon
density. A first exploratory study with respect to antikaon
selfenergies by Li, Ko and Fang in Ref.~\cite{Kolix} indicated that
sizeable attractive $K^-$ potentials are needed to explain the experimental
spectra from~\cite{Schro} for Ni~+~Ni at 1.85~A$\cdot$GeV . Indeed, their
findings could be substantiated recently in Ref.~\cite{WC96} in a
systematic analysis of antikaon production in nucleus-nucleus
collisions. So far the production of $K^+$ mesons has been primarily
addressed in order to learn about the incompressibility of the nuclear
equation of state \cite{Aich,Tomo}.  On the other hand, this process
could also give information on the predicted  repulsive kaon potential.
This should also be seen in reduced kaon production yields since at
1 -- 2~A$\cdot$GeV  already considerable baryon densities (up to 3~$\rho_0
\approx 0.5 \ {\rm fm}^{-3}$) can be reached in the compression
phase~\cite{Cass}. According to the study in Ref.~\cite{waas} the
repulsive kaon potential might be even about 100~MeV at $2 \rho_0$ which
should lead to a sizeable reduction of the cross section.

The most serious problem related to $K^+$ production at 'subthreshold'
energies are the baryon-baryon elementary production cross sections
close to threshold where no experimental data had been available so far
and rough extrapolations have been used \cite{schur,randrup}. Only very
recently a new experimental data point has been obtained 2~MeV above
the kaon production threshold in $pp$ collisions \cite{COSY}, which can
be used for more accurate parametrizations of the elementary production
cross section and a systematic reanalysis of the experimental data
taken up to now.

Our paper is organized as follows: In Section 2 we briefly describe
the transport approach employed as well as the new parametrizations for
the elementary strangeness production channels. Section 3 contains a
detailed comparison of our calculations with the available data at SIS
energies while Section 4 concludes our study with a summary.

\section{Model -- Ingredients}
In this paper we perform our analysis along the line of the
HSD\footnote{Hadron String Dynamics} approach~\cite{Ehehalt} which is
based on a coupled set of covariant transport equations for the
phase-space distributions $f_{h} (x,p)$ of hadron
$h$~\cite{Ehehalt,Weber1}, i.e.
\begin{eqnarray}  \label{g24}
\lefteqn{\left\{ \left( \Pi_{\mu}-\Pi_{\nu}\partial_{\mu}^p U_{h}^{\nu}
-M_{h}^*\partial^p_{\mu} U_{h}^{S} \right)\partial_x^{\mu}
+ \left( \Pi_{\nu} \partial^x_{\mu} U^{\nu}_{h}+
M^*_{h} \partial^x_{\mu}U^{S}_{h}\right) \partial^{\mu}_p
\right\} f_{h}(x,p) } \nonumber \\
&& = \sum_{h_2 h_3 h_4\ldots} \int d2 d3 d4 \ldots
 [G^{\dagger}G]_{12\to 34\ldots}
\delta^4(\Pi +\Pi_2-\Pi_3-\Pi_4 \ldots )  \nonumber\\
&& \times \left\{ f_{h_3}(x,p_3)f_{h_4}(x,p_4)\bar{f}_{h}(x,p)
\bar{f}_{h_2}(x,p_2)\right.  \nonumber\\
&& -\left. f_{h}(x,p)f_{h_2}(x,p_2)\bar{f}_{h_3}(x,p_3)
\bar{f}_{h_4}(x,p_4) \right\} \ldots\ \ .
\end{eqnarray}
In Eq.~(\ref{g24}) $U_{h}^{S}(x,p)$ and $U_{h}^{\mu}(x,p)$ denote the
real part of the scalar and vector hadron selfenergies, respectively,
while $[G^+G]_{12\to 34\ldots} \delta^4 (\Pi
+\Pi_2-\Pi_3-\Pi_4\ldots )$ is the 'transition rate' for the process
$1+2\to 3+4+\ldots$ which is taken to be on-shell in the
semiclassical limit adopted. The hadron quasi-particle properties in
(\ref{g24}) are defined via the mass-shell constraint~\cite{Weber1},
\begin{equation}   \label{g25}
\delta (\Pi_{\mu}\Pi^{\mu}-M_{h}^{*2} ) \ \ ,
\end{equation}
with effective masses and momenta (in local Thomas-Fermi approximation)
given by
\begin{eqnarray}\label{g26}
M_{h}^* (x,p)&=&M_h + U_h^{{S}}(x,p) \nonumber \\
\Pi^{\mu} (x,p)&=&p^{\mu}-U^{\mu}_h (x,p)\ \ ,
\end{eqnarray}
while the phase-space factors
\begin{equation}
\bar{f}_{h} (x,p)=1 \pm f_{{h}} (x,p)
\end{equation}
are responsible for fermion Pauli-blocking or Bose enhancement,
respectively, depending on the type of hadron in the final/initial
channel. The dots in Eq.~(\ref{g24}) stand for further contributions to
the collision term with more than two hadrons in the final/initial
channels. The transport approach (\ref{g24}) is fully specified by
$U_{h}^{S}(x,p)$ and $U_{h}^{\mu}(x,p)$ $(\mu =0,1,2,3)$, which
determine the mean-field propagation of the hadrons, and by the
transition rates $G^\dagger G\,\delta^4 (\ldots )$ in the collision
term, that describes the scattering and hadron production/absorption
rates.

The scalar and vector mean fields $U_{h}^{S}$ and $U^\mu_{h}$ for
baryons are taken from Ref.~\cite{Ehehalt}; the hyperon selfenergies
are assumed to be $2/3$ of the nucleon selfenergies as in
\cite{Ehehalt}.  In the present approach we propagate explicitly pions,
$\eta$'s, $\rho$'s, $\omega$'s and $\Phi$'s as free particles whereas kaons
and antikaons are propagated with effective potentials.

As in case of antiprotons there are several models for the kaon and
antikaon selfenergies \cite{GB1}--\cite{waas}, which all differ in the
actual magnitude of the selfenergies, but agree on the relative signs
for kaons and antikaons. Thus in line with the kaon-nucleon scattering
amplitude the $K^+$ potential should be slightly repulsive at finite
baryon density whereas the antikaon should see an attractive potential
in the nuclear medium. Without going into a detailed discussion of the
various models we adopt the more practical point of view, that the
actual $K^+$ selfenergies are unknown and as a guide for our analysis
use a linear extrapolation of the form (cf. \cite{WC96}),
\begin{equation}
\label{kmass}
m^*_K(\rho_B) = m_K^0 \left(1 + \alpha \frac{\rho_B}{\rho_0}\right),
\end{equation}
with $\alpha$ describing the strength of the kaon potential at finite baryon
density $\rho_B$. For our following analysis we will restrict to $\alpha$ = 0
and $\alpha$ = 0.06 to model a slightly repulsive kaon potential. We note
that the kaon potential determined from the kaon-nucleon scattering
length $a_{KN}$ within the impulse approximation \cite{impulse} leads to
$\alpha \approx$ 0.06 when using the isospin averaged scattering length
$\bar{a}_{KN} \approx - 0.255$~fm from Ref.~\cite{Barnes}.

First, the individual production channels of the kaons have to be
specified. Here, we express the cross sections as a function of the
scaled invariant energy squared $s_0/s$, since the change of the
quasi-particle mass then can be incorporated in the threshold energy
$\sqrt{s_0}$ for the particular channel. This recipe might be still a
matter of debate;  our findings in
Refs.~\cite{Sibirt1,Sibirt2,Lyk96,Sibirt3} indicate, that the
production is essentially dominated by phase space close to threshold
and thus a scaling in $s_0/s$ should be a good approximation.

The isospin averaged  production cross section of a $K^+ \Lambda$ and
$K^+ \Sigma$ pair in a nucleon-nucleon collision is related to the
measured isospin channels as:
\begin{eqnarray}
&&\sigma_{NN \to K^+ \Lambda N} = {3\over 2} \sigma_{pp \to K^+ \Lambda p}
 \label{nnkp}\\
&&\sigma_{NN \to K^+ \Sigma N} = {3\over 2} \left(
\sigma_{pp \to K^0 \Sigma^+ p} + \sigma_{pp \to K^+ \Sigma^0 p} \right).
 \label{nnkp0}
\end{eqnarray}
Following~\cite{Sibirt2} the reaction cross section can be
approximated by
\begin{eqnarray}
&& \sigma_{pp \to K^+ \Lambda p}(s) = 732 \
{\left( 1 - \frac {s_{01}} {s} \right) }^{1.8}
{\left(\frac {s_{01}} {s} \right) }^{1.5} \ \ [{\mu}b]
  \label{Sppkp1}\\
&& \sigma_{pp \to K^0 \Sigma^+ p}(s) = 338.46 \
{\left( 1 - \frac {s_{02}} {s} \right) }^{2.25}
{\left(\frac {s_{02}} {s} \right) }^{1.35} \ \ [{\mu}b]
  \label{Sppkp2}\\
&& \sigma_{pp \to  K^+ \Sigma^0 p}(s) = 275.27 \
{\left( 1 - \frac {s_{02}} {s} \right) }^{1.98}
{\left(\frac {s_{02}} {s} \right) } \ \ [{\mu}b]
  \label{Sppkp3}
\end{eqnarray}
with $\sqrt{s_{01}} = m_\Lambda - m_N + m^0_K$ and $\sqrt{s_{02}} =
m_\Sigma + m_N + m^0_K$. According to isospin relations the $N\Delta$
and $\Delta\Delta$ production channels get additional factors of 3/4
and 1/2, respectively. This scaling of the $\Delta N$ and $\Delta \Delta$
production channels from Ref. \cite{randrup}, however, is questionable
and further microscopic studies along the lines of
Refs.~\cite{LiKo,Peters}  will be necessary to determine these reaction
cross sections more accurately.

The elementary cross sections for the pion
induced channels $\pi N \to K^+ Y$ have been computed by Tsushima et
al. in Ref.~\cite{Ts1}, which  we adopt for our present study.  $K^+$
elastic scattering with nucleons also has an impact on the final kaon
spectra; the elastic cross section employed is displayed in Fig.~5 of
Ref.~\cite{WC96}.  In addition, we include the $K\bar{K}$ production
channels in baryon-baryon ($BB$), $\pi B$ and $\pi \pi$ collisions
within the parametrizations from \cite{WC96} which, however, do not
contribute significantly to the inclusive $K^+$ yield.

The calculation of 'subthreshold' particle production is described in
detail in Refs.~\cite{Cass,Cass90} and has to be treated perturbatively
in the energy regime of interest here due to the small cross sections
involved. Since we work within the parallel ensemble algorithm, each
parallel run of the transport calculation can be considered
approximately as an individual reaction event, where binary reactions
in the entrance channel at given invariant energy $\sqrt{s}$ lead to
final states with 2 (e.g. $K^+ Y$ in $\pi B$ channels) or 3 (e.g. for
$K^+ YN$ channels in $BB$ collisions) particles with a relative weight
$W_i$ for each event $i$ which is defined by the ratio of the
production cross section to the total hadron-hadron cross
section\footnote{The actual final states are chosen by Monte Carlo sampling
according to the 2, or 3-body phase space.}.  The perturbative
treatment now implies that in case of strangeness production channels
the initial hadrons are not modified in the respective final channel.
On the other hand, each strange hadron is represented by a testparticle
with weight $W_i$ and propagated according to the Hamilton equations of
motion. Elastic and inelastic reactions with pions, $\eta$'s or
nonstrange baryons are computed in the standard way~\cite{Cass};
however, only the dynamical feedback of the strange hadrons to the
nonstrange mesons and baryons is neglected.  The final cross section is
obtained by multiplying each testparticle with its weight $W_i$.  In
this way one achieves a time-saving simulation of the strangeness
production, propagation and reabsorption during the heavy-ion
collision.

\section{Results}
\begin{figure}[h1]
\vspace*{-2cm}
{\psfig{figure=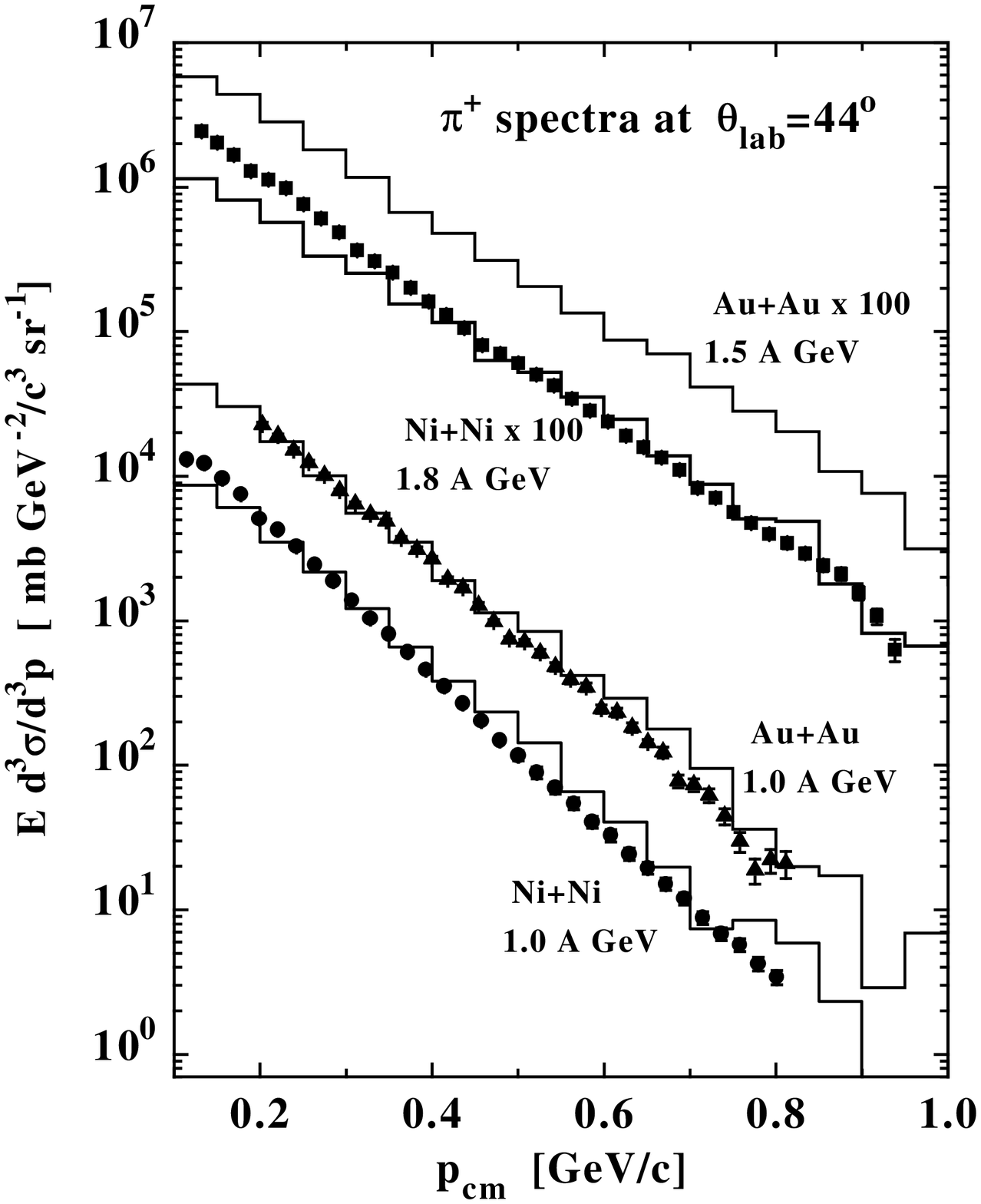,width=15cm,height=20cm}}
\vspace*{-2cm}
\caption{Inclusive $\pi^+$ spectra from Ni~+~Ni collisions at 1.0 and
1.8~A$\cdot$GeV  and Au~+~Au at 1.0~A$\cdot$GeV  in comparison to the
experimental
data from Ref.~\protect\cite{expion} at $\theta_{lab} = 44^0 \pm 4^o$.
The upper solid histogram displays our calculations for Au~+~Au at
1.5~A$\cdot$GeV . }
\label{Fig1}
\end{figure}

Since $\Delta$- and pion-induced channels play a major role for $K^+$
production, we start our analysis with a comparison of our calculations
for pion production at SIS energies with the available $\pi^+$ data for
Ni~+~Ni at 1.0 and 1.8~A$\cdot$GeV  and Au~+~Au at 1.0~A$\cdot$GeV  in
Fig.~\ref{Fig1}.  The experimental spectra at $\theta_{lab} = 44^o \pm
4^o$ \cite{expion} are described reasonably well in the whole
kinematical range not only for Ni~+~Ni, but also for Au~+~Au. We
slightly underestimate the pion spectrum for Ni + Ni at 1.8 A$\cdot$GeV  at
low momenta which might reflect limitations of our configuration space
showing up at higher bombarding energy.  For future comparison we also
include our results for the inclusive $\pi^+$ spectra from Au + Au at
1.5 A$\cdot$GeV  (upper histogram). For a more detailed analysis of pion
production at SIS energies we refer the reader to Ref.~\cite{Teis96},
which works in a very similar theoretical framework.  For our present
purpose we only conclude that the pion and baryon dynamics including
the $\Delta$ and higher resonance dynamics in our transport
calculations are sufficiently well under control.

\begin{figure}[h]
\vspace*{-2cm}
{\psfig{figure=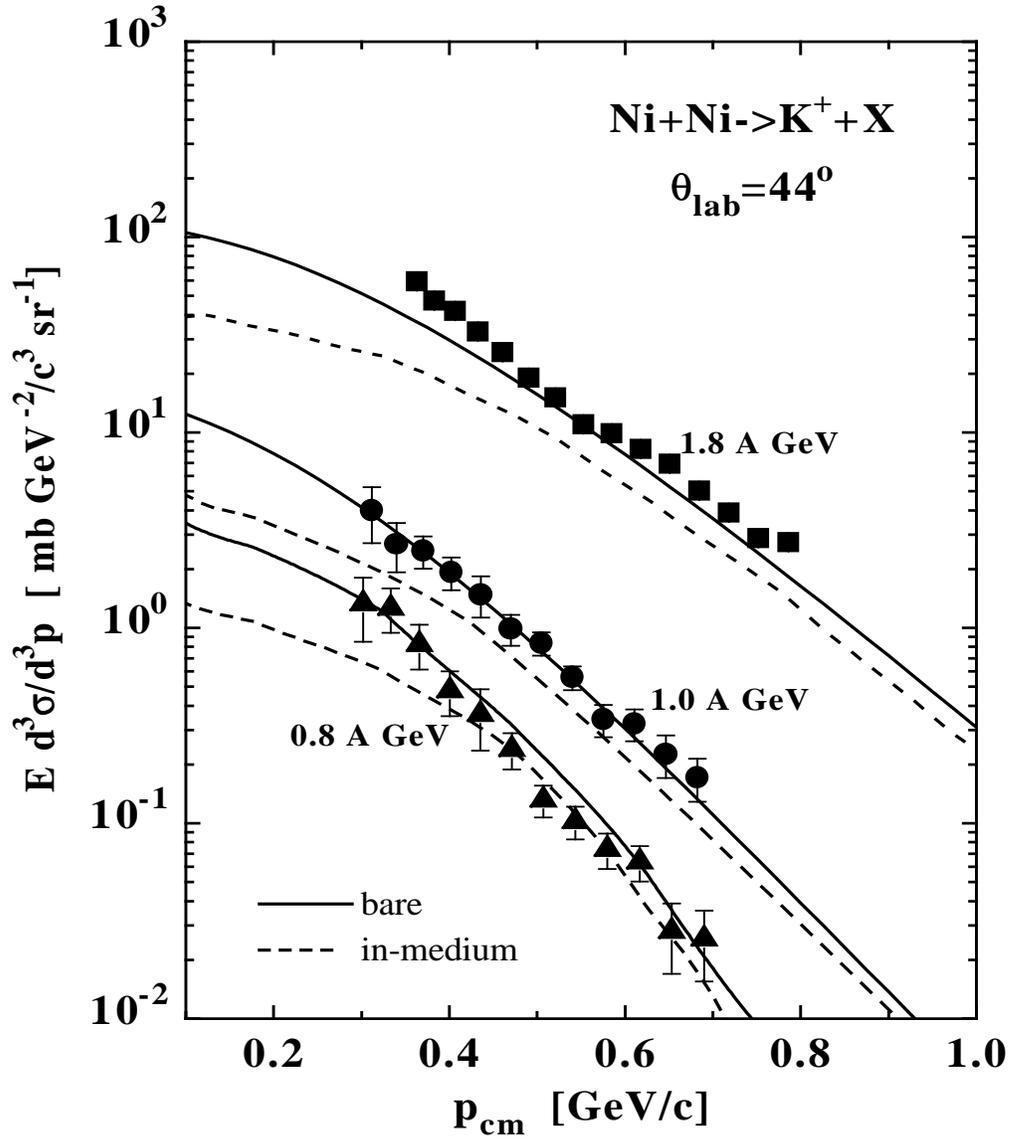,width=15cm,height=20cm}}
\vspace*{-2cm}
\caption{Inclusive $K^+$ spectra from Ni~+~Ni collisions at 0.8, 1.0
and 1.8~A$\cdot$GeV  in comparison to the experimental data from
Ref.~\protect\cite{Senger} at $\theta_{lab} = 44^0 \pm 4^o$. The solid
lines represent calculations with bare $K^+$ masses, while the dashed
lines result for $\alpha$ = 0.06 in Eq.~(\protect\ref{kmass}).}
\label{Fig2}
\end{figure}

The Lorentz invariant $K^+$ spectra for Ni~+~Ni at 0.8, 1.0 and
1.8~A$\cdot$GeV  from the same calculations as above are shown in
Fig.~\ref{Fig2} in
comparison to the data from \cite{Senger}. Here the full lines reflect
calculations including only bare $K^+$ masses ($\alpha$ = 0) while the
dashed lines correspond to calculations with $\alpha$ = 0.06 in
Eq.~(\ref{kmass}), which leads to about a 30 MeV increase of the kaon
mass at $\rho_0$.  Note that the repulsive kaon potential from
Ref.~\cite{waas} is best fitted for $\alpha$ = 0.1, which would imply a
further reduction of the $K^+$ cross section. When fitting an exponential
to the Lorentz invariant spectrum $\sim \exp(- E/T_0)$, where $E$ is the
kinetic energy of the kaon in the cms, we obtain slope parameters $T_0
\approx $ 60 MeV, 75 MeV and 102 MeV at the bombarding energy of
0.8, 1.0 and 1.8 A$\cdot$GeV , respectively.

\begin{figure}[h]
\vspace*{-2cm}
{\psfig{figure=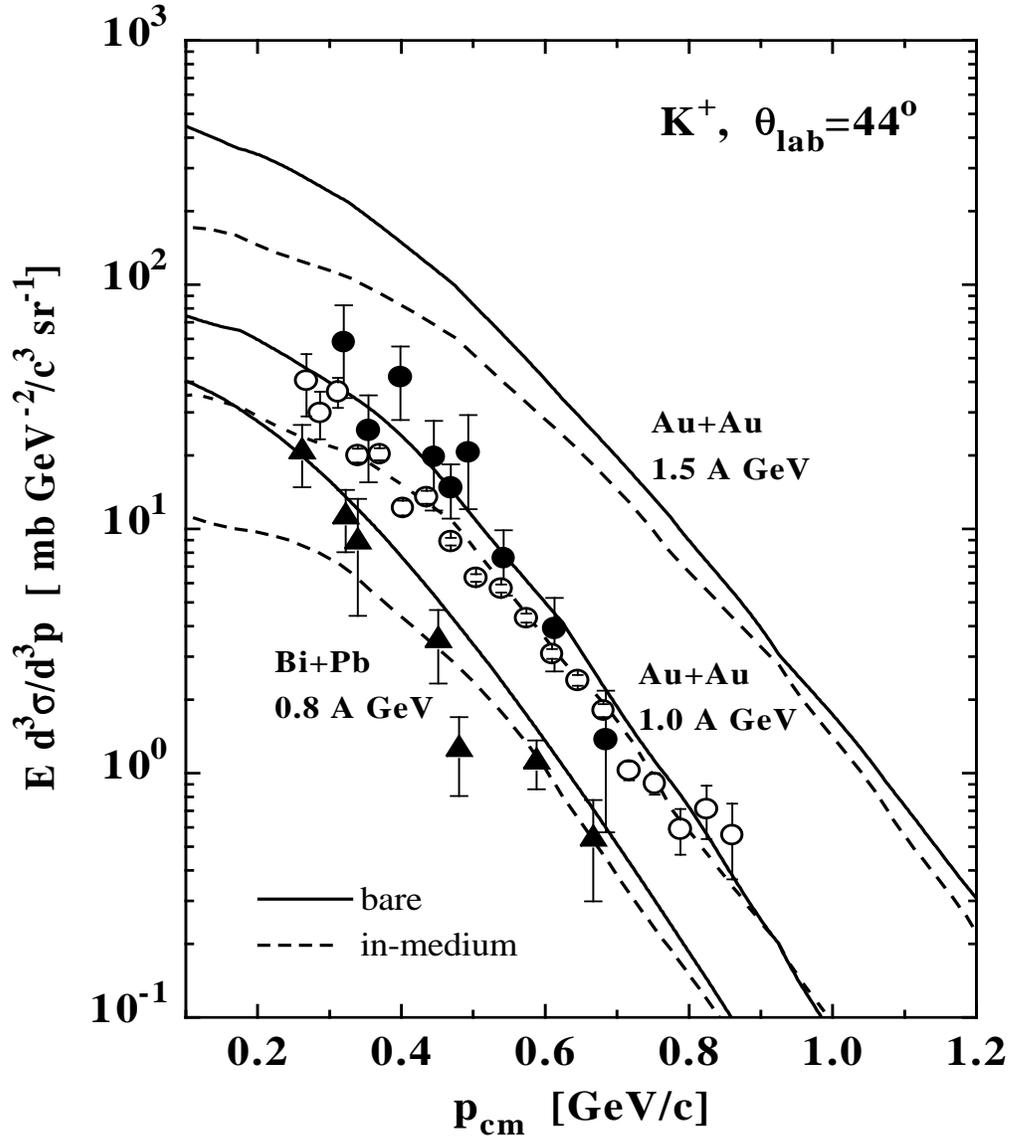,width=15cm,height=20cm}}
\vspace*{-2cm}
\caption{Inclusive $K^+$ spectra from Bi~+~Pb collisions at 0.8~A$\cdot$GeV
and Au~+~Au at 1.0~A$\cdot$GeV  in comparison to the experimental data from
Ref.~\protect\cite{BiPb,Misko,SengerAu} at $\theta_{lab} = 44^0 \pm
4^o$. The solid lines represent calculations with bare $K^+$ masses,
while the dashed lines result for $\alpha$ = 0.06 in
Eq.~(\protect\ref{kmass}). The calculations for Au~+~Au at 1.5~A$\cdot$GeV
are included for future comparison.}
\label{Fig3}
\end{figure}

The general tendency seen at all bombarding energies is that our
calculations with a bare kaon mass provide a better description of the
experimental data for Ni + Ni than those with an enhanced kaon mass.
Since this general trend might be accidental we also compare our
calculations for the heavier systems, i.e. Bi~+~Pb at 0.8~A$\cdot$GeV
and Au~+~Au at 1.0~A$\cdot$GeV , with the respective experimental data
\cite{BiPb,Misko,SengerAu} in Fig.~\ref{Fig3}. The full dots represent
the earlier data for Au + Au at 1.0~A$\cdot$GeV  \cite{Misko} while the
open circles result from a new measurement \cite{SengerAu} of the
system at the same bombarding energy. Both calculations (solid line:
$\alpha$ = 0; dashed line:  $\alpha$ = 0.06) are compatible with the
data due to the experimental uncertainties.  As in case of
antikaons~\cite{WC96} the selfenergy effects are most pronounced at low
center-of-mass momenta of the kaons where, unfortunately, no data have
been taken so far. For future comparison we also include our results
for Au~+~Au at 1.5~A$\cdot$GeV .

\begin{figure}[h]
\vspace*{-2cm}
{\psfig{figure=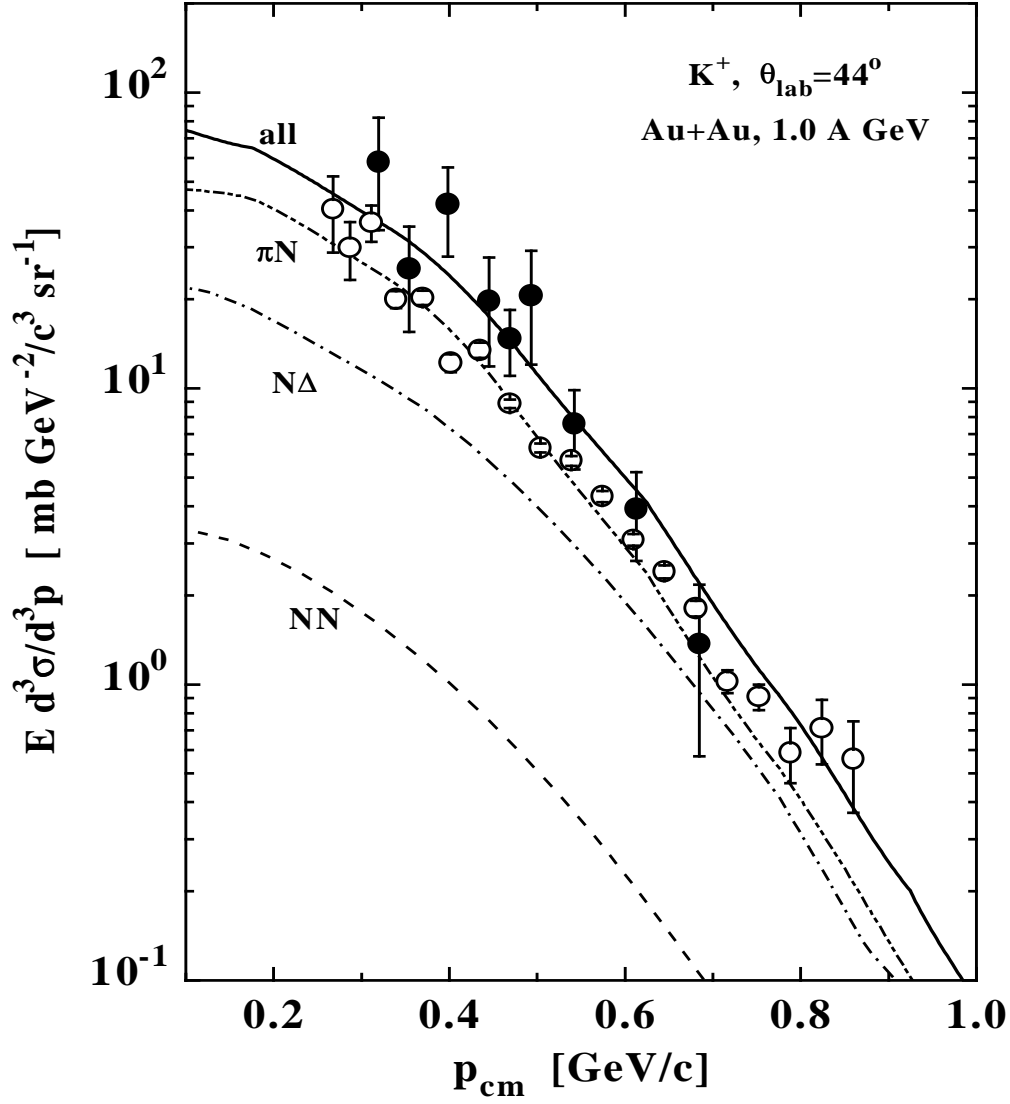,width=15cm,height=20cm}}
\vspace*{-2cm}
\caption{Inclusive $K^+$ spectra from  Au~+~Au collisions at 1.0~A$\cdot$GeV
(solid line) in comparison to the experimental data from
Ref.~\protect\cite{Misko,SengerAu} at $\theta_{lab} = 44^0 \pm 4^o$ for
the bare kaon mass scenario. The dashed line represents the
contribution from $NN$ collisions while the dot-dashed and
dot-dot-dashed line show the contributions from $\Delta N$ and $\pi N$
collisions, respectively.}
\label{Fig4}
\end{figure}

With our new elementary cross sections the relative weights of the
$K^+$ production channels change considerably compared to our earlier
calculations \cite{Lang} as can be seen from Fig.~\ref{Fig4} for Au +
Au at 1~A$\cdot$GeV . Whereas the $NN$ production channels (dashed
line) are almost negligible as before \cite{Lang} the dominant yield
now stems from $\pi N$ reactions (dot-dot-dashed line) which surpass
the $N \Delta$ channel (dot-dashed line).  This change comes about
because our old calculations \cite{Lang} were performed in the
so-called frozen-$\Delta$ approximation in which the $\Delta$'s are not
allowed to decay until the collision has ended.  The main message of
the old studies, that the secondary reactions are very important for a
description of the kaon yield, however, survives.  The calculation in
Fig.~\ref{Fig4} has been performed for a bare $K^+$ mass; we note that
the relative channel decomposition does not change very much when
performing a calculation with a slightly repulsive kaon potential.

In Fig.~\ref{Fig5} we present the ratio $N_{K^+}/N_{\pi^+}$ versus the
number of participating nucleons $A_{part}$ -- which is a measure of
the impact parameter $b$ -- from our calculations for Ni~+~Ni at 1.0
and 1.8~A$\cdot$GeV  as well as for Au~+~Au at 1.0 and 1.5~A$\cdot$GeV
for $\alpha$ = 0 (solid lines) and $\alpha$ = 0.06 (dashed lines).  The
experimental data have been taken from Ref.~\cite{Misko}. The
'theoretical' error bars indicate the calculational uncertainty due to
the finite particle statistics, since only mesons for $\theta_{lab} =
44^o \pm 4^o$ have been taken into account. The ratio
$N_{K^+}/N_{\pi^+}$ increases with $A_{part}$ in all cases; this
increase is more pronounced at the lower bombarding energy of
1.0~A$\cdot$GeV  and for the heavier system Au~+~Au.  In our
calculation this increase in the relative $K^+$ yield is due to a
strong rise in $N_{K^+}/A_{part}$; $N_{\pi}/A_{part}$ is flat as a
function of $A_{part}$.  Because the kaons are mainly due to the
secondary collisions, they are predominantly produced at high baryon
density $1.5 \rho_0 < \rho_B < 2.5 \rho_0$ while the pion production
occurs at lower density, too.  Since the volume for the high baryon
density increases sizeably when going from peripheral to more central
collisions, the $K^+/\pi^+$ ratio has to increase accordingly.  The
relative contribution of the various production channels to the total
$K^+$ yield are shown in Fig.~\ref{Fig6} as a function of $A_{part}$
for Ni + Ni at 1.0 and 1.8~A$\cdot$GeV . It can clearly be seen that
the relative importance of the primary $NN$ channel increases with
energy, but that of the secondary $\pi N$ reaction decreases. This
explains the flatter behavior seen in Fig.~\ref{Fig5} for the higher
energies.

In Fig.~\ref{Fig7} we show the angular distribution of the produced
kaons for Au~+~Au at 1~A$\cdot$GeV  which exhibits a forward-backward
peaking.  By comparing our calculations with and without angular
dependence in the elementary kaon production cross section  we have
verified that this angular anisotropy is not due to the elementary
production processes.  Instead, it reflects the forward-backward
peaking of the pion angular distribution \cite{Teis96}. The figure
shows that the rescattering of the kaons enhances the anisotropy
somewhat, but is not solely responsible for it.

A further quantity of interest -- in the context of our present
analysis -- is the kaon flow in the reaction plane, which should show
some sensitivity to the kaon potential in the nuclear medium as put
forward by Li, Ko and Brown \cite{flow1,flow2,flow3,flow4} and is
presently investigated by the FOPI Collaboration \cite{flow5,Norbert}.
Here due to elastic scattering with nucleons the kaons partly flow in
the direction of the nucleons thus showing a positive flow in case of
no mean-field potentials \cite{flow1}.  With increasing repulsive kaon
potential the positive flow will turn to zero and then become negative;
experimental data on kaon flow thus are expected to discriminate
further on the potentials seen in the medium. In order to investigate
this question we have performed detailed (and high statistics)
calculations for $K^+$ production in Ni + Ni reactions at 1.93 A$\cdot$GeV
as measured by the FOPI Collaboration \cite{Norbert}. In order to compare
with their data we have included a transverse momentum cut $p_T \geq$
0.5 $m_K$, where $m_K$ is the kaon mass, and gated on central collisions
with impact parameter $b \leq$ 4 fm as in Ref. \cite{flow3}.  The
results of our calculations are displayed in Fig. \ref{Fig8} in terms
of $<p_x>/m_K$ versus the normalized rapidity
\begin{equation}
y^0 = \frac{y_{cm}}{y_{proj}},
\label{yflow}
\end{equation}
where $y_{cm}$ and $y_{proj}$ are the kaon and projectile rapidity
in the cms, respectively.  Our calculations without any kaon potential
($\alpha$ = 0, full line) indeed show a positive flow as expected,
which appears still to be compatible with the data within the error
bars and is practically identical to the respective calculations from
Ref.  \cite{flow3}. On the other hand, increasing the kaon potential
($\alpha$ = 0.06, dashed line) the kaon flow becomes slightly negative
or comparable to zero, quantitatively in line with the calculations
from Ref. \cite{flow3}  and in somewhat better agreement with the
data. Thus in case of the flow observable a slightly repulsive kaon
potential ($\approx$ 30 MeV at $\rho_0$) is more in line with the
present data of the FOPI Collaboration \cite{Norbert}.

\section{Summary}
In this work we have presented a detailed study of pion and kaon
production in nucleus-nucleus collisions for medium and heavy systems
from 0.8 to 1.8~A$\cdot$GeV  within the coupled channel BUU approach,
where the kaons are produced perturbatively, however, propagated
explicitly with their final state interactions. An important ingredient
of our reanalysis of the kaon cross sections are the novel elementary
production cross sections from Refs.~\cite{Sibirt1,Sibirt2,Sibirt3}
and from Ref.~\cite{Ts1} for the pion induced channels. We note,
however, that the $\Delta N$ and $\Delta \Delta$ production channels
are not that well determined and require further theoretical efforts.

\begin{figure}[h]
\vspace*{-2cm}
{\psfig{figure=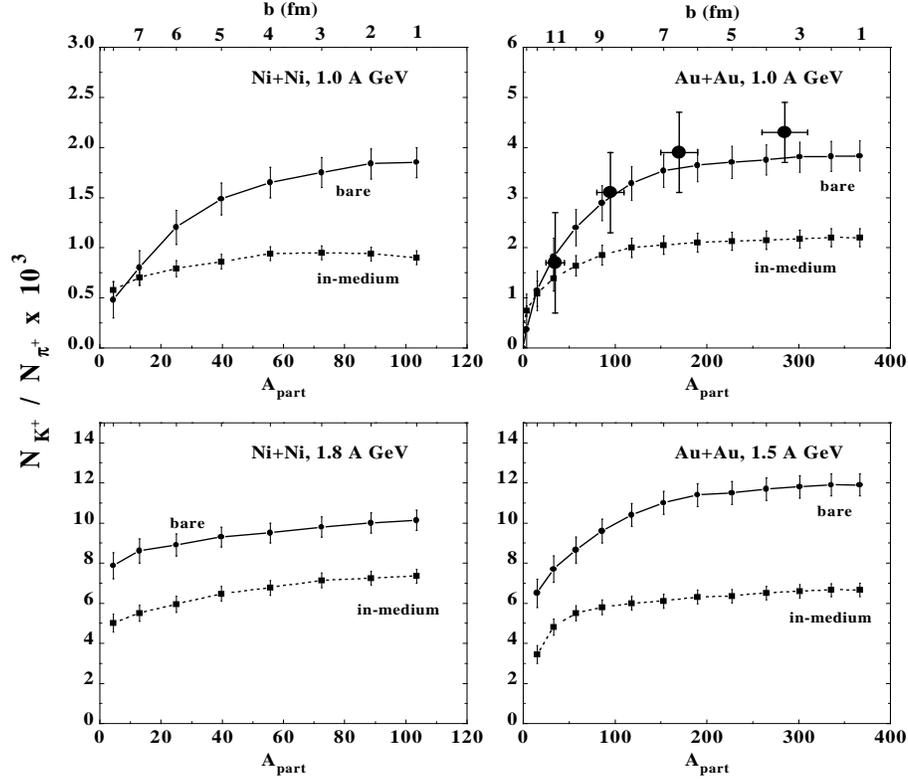,width=15cm,height=20cm}}
\vspace*{-2cm}
\caption{The $K^+/\pi^+$ ratio for Ni~+~Ni at 1.0 and 1.8~A$\cdot$GeV  and
Au~+~Au at 1.0 and 1.5~A$\cdot$GeV  as a function of the number of
participating nucleons $A_{part}$ as defined in
Ref.~\protect\cite{Misko}.  The solid lines represent
calculations with bare $K^+$ masses, while the dashed lines result for
$\alpha$ = 0.06 in Eq.~(\protect\ref{kmass}).  The experimental data
have been taken from Ref.~\protect\cite{Misko}.}
\label{Fig5}
\end{figure}

\begin{figure}[h]
\vspace*{-2cm}
{\psfig{figure=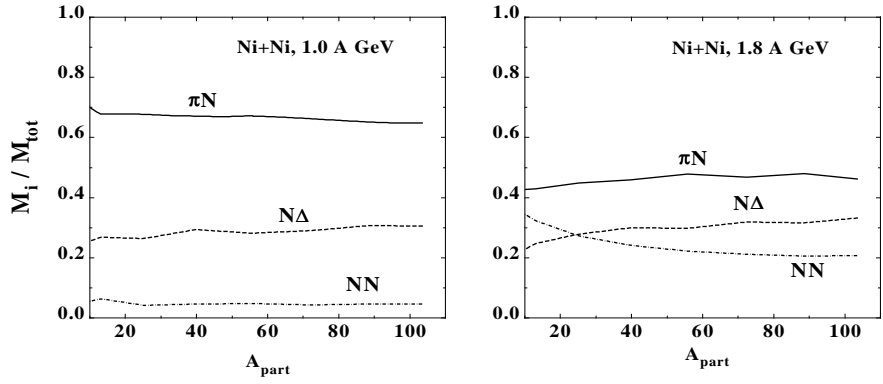,width=15cm,height=20cm}}
\vspace*{-2cm}
\caption{The relative contribution of the various production channels
to the total $K^+$ yield as a function of $A_{part}$ for Ni + Ni at 1.0
and 1.8~A$\cdot$GeV .  The solid line represents the contribution from $\pi N$
collisions while the dashed and dot-dashed lines show the contributions
from $\Delta N$ and $NN$ collisions, respectively.}
\label{Fig6}
\end{figure}

\begin{figure}[h]
\vspace*{-2cm}
{\psfig{figure=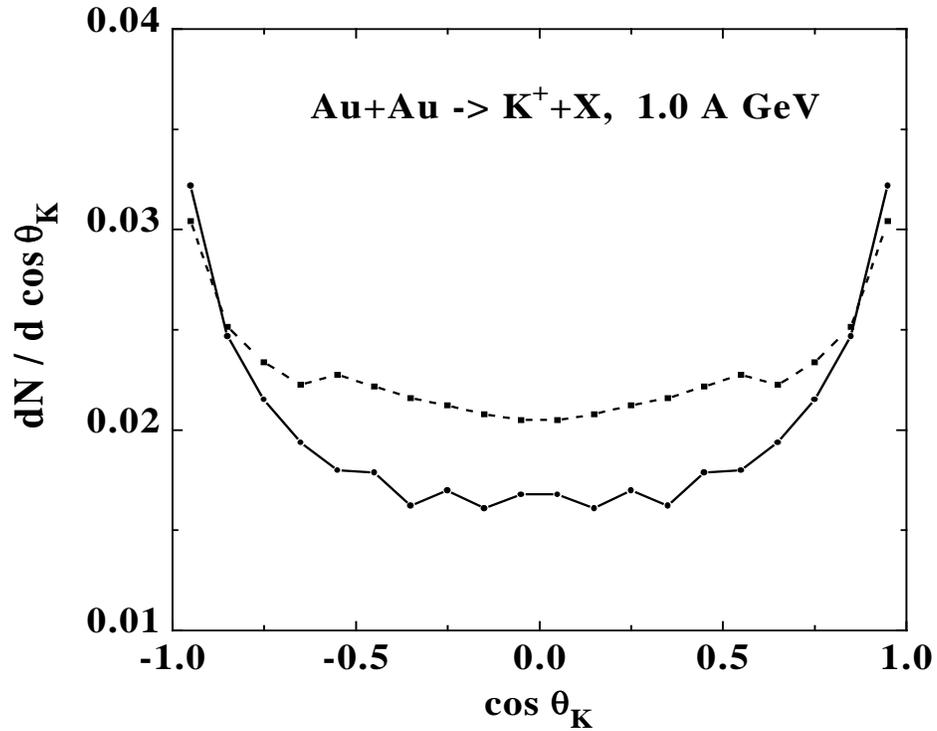,width=15cm,height=20cm}}
\vspace*{-2cm}
\caption{Angular distribution of kaons in the nucleus-nucleus cms
for Au~+~Au collisions at 1~A$\cdot$GeV  including all impact parameters.
The dashed line represents a calculation without $K^+N$ rescattering
while the solid line shows the results when including kaon rescattering.}
\label{Fig7}
\end{figure}

\begin{figure}[th1]
\vspace*{-2cm}
{\psfig{figure=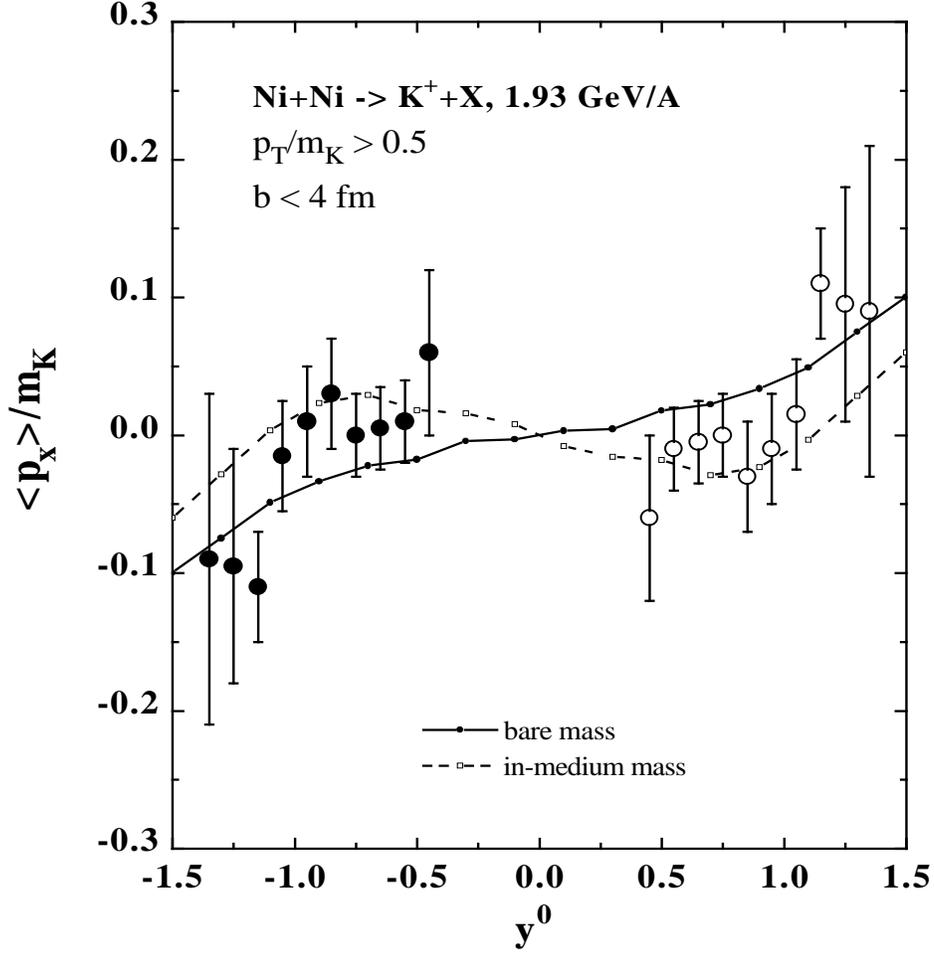,width=15cm,height=20cm}}
\vspace*{-2cm}
\caption{Kaon flow in the reaction plane ($<p_x>/m_K)$ as a function of
the normalized rapidity $y^0 = y_{cm}/y_{proj}$  for Ni + Ni
at 1.93 A GeV. We have gated on central collisions ($b \leq$ 4 fm) and
applied a transverse momentum cut for $p_T \geq$ 0.5 $m_K$ as for the
experimental data of the FOPI Collaboration \protect\cite{Norbert}
(full dots). The open dots are obtained by reflection at $y^0$ = 0. The
solid line and dashed line display the results of our transport
calculations without ($\alpha=0$) and with a slightly repulsive
kaon potential ($\alpha=0.06$), respectively.}
\label{Fig8}
\end{figure}

Our analysis shows that the $\pi^+$ spectra are reasonably  well
described in this energy regime without introducing any medium
modifications for these mesons (cf. also Ref.~\cite{Teis96}).  This is
understood in terms of the strong reabsorption of pions that
essentially leads to the surface-emission.  For the much more
penetrating $K^+$ mesons our results also show no final convincing
indications of in-medium changes.  The new kaon spectra for the heavy
system Au + Au at 1 A$\cdot$GeV as well as the kaon flow data for Ni +
Ni at 1.93 A$\cdot$GeV  can -- within their errorbars -- be
described by assuming free kaon properties although there seems to be a
tendency to favor a small repulsive kaon potential of about +30 MeV at
normal nuclear matter density, which would be in line with the kaon
potential as extracted from the kaon-nucleon scattering length in the
impulse approximation \cite{impulse}. On the other hand, we find that
more repulsive kaon potentials as predicted by some Lagrangian models
appear not to be compatible with the available data and in particular
with the $K^+/\pi^+$ ratios. An upcoming data analysis in the form of
Fig.~\ref{Fig5} and high statistics data on kaon flow in the form of
Fig. \ref{Fig8} should shed some further light on this issue.
However, even then theoretical uncertainties connected with our
insufficient knowledge of $\Delta$-induced cross sections will remain.

\vspace{4mm}
The authors acknowledge valuable and inspiring discussions throughout
this work with N.~Herrmann, H.~Oeschler, P.~Senger, A.~Sibirtsev and
Gy.~Wolf.

\newpage

\end{document}